\documentclass[a4paper,12pt]{article}
\usepackage{graphics}
\usepackage{amsmath,amssymb}
\usepackage{color}
\def\vf{\varphi}
\def\ve{\varepsilon}
\def\half{\frac{1}{2}}

\def\bs{\boldsymbol}
\def\be{\begin{equation}}
\def\ee{\end{equation}}
\def\bea{\begin{eqnarray}}
\def\eea{\end{eqnarray}}
\def\beax{\begin{eqnarray*}}
\def\eeax{\end{eqnarray*}}
\begin{document}
\setcounter{footnote}{1}
\renewcommand{\thefootnote}{{\fnsymbol{footnote}}}
\title{Gauging non-Hermitian Hamiltonians}
\author{H. F. Jones\footnote{e-mail: h.f.jones@imperial.ac.uk}\\
Physics Department, Imperial College, London SW7 2AZ, UK}
\date{}
\maketitle
\begin{abstract}
We address the problem of coupling non-Hermitian systems, treated as fundamental
rather than effective theories, to the electromagnetic field. In such theories
the observables are not the $\bs{x}$ and $\bs{p}$ appearing in the Hamiltonian, but quantities
$\bs{X}$ and $\bs{P}$ constructed by means of the metric operator. Following the analogous
procedure of gauging a global symmetry in Hermitian quantum mechanics we find that the
corresponding gauge transformation in $\bs{X}$ implies minimal substitution in the form $\bs{P}\to \bs{P} - e\bs{A}(\bs{X})$.
We discuss how the relevant matrix elements governing electromagnetic transitions may be calculated
in the special case of the Swanson Hamiltonian, where the equivalent Hermitian Hamiltonian
$h$ is local, and in the more generic example of the imaginary cubic interaction, where $H$ is local
but $h$ is not.\\

\noindent PACS numbers: 03.65.Ca, 11.30.Er, 02.30.Mv
\end{abstract}

\section{Introduction}
Recent interest in Hamiltonians that are non-Hermitian but nonetheless have a real spectrum dates from
the pioneering paper of Bender and Boettcher\cite{BB}, which gave strong numerical and analytical evidence
that the spectrum of the class of Hamiltonians
\bea
H=p^2+m^2x^2-(ix)^N
\eea
was completely real and positive for $N\ge 2$, and attributed this reality to the (unbroken) $PT$ symmetry of the Hamiltonian.
Subsequently a large number of $PT$-symmetric models were explored (see, e.g. \cite{GZ}), and it was found that the phenomenon
was rather general. The drawback that the natural metric on the Hilbert space, with
overlap $\int \psi_i(-x)\psi_j(x)dx$, was not positive definite, was overcome by the realization\cite{BBJ} that
one could construct an alternative, positive-definite metric involving the so-called $C$ operator.
The formalism was further developed by Mostafazadeh\cite{AM-metric},  building on earlier work by Scholtz et al.\cite{Hendrik}.
In particular he showed\cite{AM-h} that such a Hamiltonian $H$ was related by a similarity transformation to an equivalent
Hermitian Hamiltonian $h$. The key relation is the quasi-Hermiticity of $H$:
\bea\label{QH}
H^\dag=\eta H \eta^{-1},
\eea
where $\eta$ is Hermitian and positive definite. $\eta$ is related to the $C$ operator by
$\eta=CP$, and it is frequently extremely useful\cite{Q} to write it in the exponential form $\eta=e^{-Q}$.
Occasionally $\eta$ can be constructed exactly (see, for example \cite{Swanson, Swanson-Geyer, Swanson-hfj,-x4, AF}), but more typically it can only be constructed in perturbation theory, for example for the $ix^3$ model\cite{BBJ-ix3}.

From Eq.~(\ref{QH}) we can immediately deduce that
\bea
h\equiv\rho H \rho^{-1},
\eea
is Hermitian, where $\rho=e^{-\half Q}$.
Other operators $A$ will also be observables, having real eigenvalues, if they are also quasi-Hermitian, i.e.
\bea
A^\dag=\eta A \eta^{-1},
\eea
and they again are related by the similarity transformation to Hermitian counterparts $a$:
\bea
A&=&\rho^{-1} a \rho\ .
\eea

The similarity transformation also transforms the states of the Hermitian system,
$|\vf\rangle$, to those of the quasi-Hermitian system, $|\psi\rangle$:
\bea
|\psi\rangle=\rho^{-1}|\vf\rangle\ .
\eea
This implies that the matrix element of an operator is
\bea
\langle {\cal O} \rangle_{ij}=\langle\psi_i|\eta\,{\cal O}|\psi_j\rangle\ .
\eea
In particular, the matrix elements of an observable can be written as
\bea
\langle\psi_i|\eta A|\psi_j\rangle&=&\langle\vf_i|\rho^{-1}\eta(\rho^{-1} a \rho)\rho^{-1}|\vf_j\rangle\nonumber\\
&=&\langle\vf_i|a|\vf_j\rangle\ .
\eea
A very important observation is that
\bea\label{H equivalence}
H(\bs{x},\bs{p})&=&H(\rho\bs{X}\rho^{-1},\rho\bs{P}\rho^{-1})\nonumber\\
&=&\rho H(\bs{X},\bs{P})\rho^{-1}\nonumber\\
&=&h(\bs{X},\bs{P}).
\eea
Thus an alternative way of finding $h$ is to calculate the observables $\bs{X}$ and $\bs{P}$ and then
rewrite $H(\bs{x}, \bs{p})$ in terms of them.

The above concerns quasi-Hermitian systems considered in isolation. However, important conceptual issues
arise when one attempts to consider such systems in interaction with an otherwise Hermitian environment.
For example, Ref.~\cite{QB} examined a non-Hermitian analogue of the Stern-Gerlach experiment in which the role
of the intermediate inhomogeneous magnetic field flipping the spin is taken over by an apparatus described by a
non-Hermitian Hamiltonian. This type of set-up has been further discussed and elaborated in a series of papers
by various authors\cite{Andreas, Uwe,AM-QB,comment,Nesterov,Uwe2}.

Again, scattering gives rise to problems, since unitarity, as conventionally defined, is generically not
satisfied for a $PT$-symmetric Hamiltonian. Unitarity can be restored, by use of the $\eta$ metric, but then
the concept of ``in" and ``out" states has to be drastically Ref.~\cite{HFJ-1df, HFJ-3df}, or in some
cases\cite{MZ-scatt} less drastically, revised.

The present paper is concerned with another such issue, namely how one couples a charged particle described
by a quasi-Hermitian Hamiltonian to the electromagnetic field, following as closely as possible the well-known
gauging procedure for a Hermitian Hamiltonian. This problem has been previously dealt with by Fariah and Fring\cite{FF}
in a treatment which in many ways is more sophisticated than the present paper, dealing with pulses rather than
plane waves and going beyond first-order perturbation theory. However, the subtleties arising from the
difference between $\bs{x}$ and $\bs{X}$ (see Eq.~(\ref{GT}) below) were not encountered there because the calculations
were done entirely within the framework of the dipole approximation, where the electromagnetic
potential $\bs{A}$ is just a function of time.

\section{Brief review of the standard procedure}
In standard quantum mechanics the probability density is just $|\psi(\bs{x})|^2$, which is
unchanged under a change of phase of the wavefunction: $\psi\to e^{ie\alpha}\psi$
provided that $\alpha$ is a real constant. If we try to extend this to $\alpha=\alpha(\bs{x})$,
a real function of $\bs{x}$, an extra term appears
in the Schr\"odinger equation, because  now $\hat{\bs{p}}\,e^{ie\alpha}\psi=e^{ie\alpha}
(\hat{\bs{p}}+e\nabla\alpha)\psi$. We cancel this additional $\nabla\alpha$ term by
\textit{minimal substitution}:
\bea
\bs{p}\to \bs{p}-e\bs{A}\ .
\eea
Then under the combined transformations
\bea\label{GT}
\left\{\begin{array}{l}\  \psi\to \psi'=e^{ie\alpha}\psi\\
\bs{A}\to \bs{A}'=\bs{A}-\nabla\alpha\ ,\end{array}\right.
\eea
we obtain ($\hat{\bs{p}}-e\bs{A})\psi\to e^{ie\alpha}(\hat{\bs{p}}-e\bs{A})\psi$, as required.
Moreover the electric and magnetic fields are unchanged by the gauge transformation (\ref{GT}).

So for a normal Hamiltonian of the form
\bea\label{Hstandardform}
H=\frac{\bs{p}^2}{2m}+V(\bs{x}),
\eea
the coupling to the vector potential is $-e(\bs{A}.\bs{p}+\bs{p}.\bs{A})/(2m)$.
In first-order perturbation theory a standard procedure then gives the transition rate between the states $|i\rangle$ and $|j\rangle$
induced by a plane wave
\bea\label{FT}
\bs{A}(\bs{x},t)=\int d\omega \tilde{\bs{A}}(\omega) e^{i(\bs{k.x}-\omega t)} + c.c.
\eea
as
\bea
w_{ij}\propto\frac{e^2}{m^2}|\langle i |p_A|j \rangle|^2,
\eea
in the dipole approximation $e^{i\bs{k.x}}\approx 1$ over the range of the interaction. Here
the constant of proportionality is $(2\pi/\hbar^2)\tilde{\bs{A}}(\omega_{ij})^2$, where $\omega_{ij}=(E_i-E_j)/\hbar$, and $p_A$ is the projection of $\bs{p}$ in the direction of $\bs{A}$.

The matrix element $\langle i |p_A|j \rangle$  can be recast in terms of $\langle i |x_A|j \rangle$, where $x_A$ is similarly defined, by
\bea
(E_i-E_j)\langle i |\bs{x}|j \rangle = \langle i |[H,\bs{x}]|j \rangle
=-\frac{i\hbar}{m}\langle i |\bs{p}\,|j \rangle,
\eea
so that
\bea
\langle i |p_A\,|j \rangle=im\omega_{ij}\langle i |x_A\,|j \rangle\ .
\eea

\section{Quasi-Hermitian quantum mechanics}
\setcounter{footnote}{1}
\renewcommand{\thefootnote}{{\fnsymbol{footnote}}}
The total\footnote{Note that the probability density
$\varrho(\bs{x})=\langle\psi|\rho|\bs{x}\rangle\langle \bs{x}|\rho|\psi\rangle$ is also invariant under the transformation
of Eq.~(\ref{inv}).} probability
is now $\langle\psi|\eta|\psi\rangle$, where $\eta$ is the metric operator.
This is no longer invariant under $|\psi\rangle\to e^{ie\alpha(\bs{x})}|\psi\rangle$, except in the special case where
$\eta=\eta(\bs{x})$, so that $[\eta,\bs{x}]=0$.

It is, however, invariant under
\bea\label{inv}
 |\psi\rangle\to e^{ie\alpha(\bs{X})}|\psi\rangle,
 \eea
where $\bs{X}$ is the observable $\bs{X}=\rho^{-1} \bs{x} \rho$. For then
\bea
\langle\psi|\eta|\psi\rangle&\to&\langle\psi| e^{-ie\alpha(\bs{X})^\dag}\eta e^{ie\alpha(\bs{X})}|\psi\rangle\nonumber\\
&=&\langle\psi|\eta|\psi\rangle,
\eea
since $\bs{X}^\dag\eta=\eta\bs{X}$.
Note that, in terms of the eigenstates $|\vf\rangle$ of $h$, Eq.~(\ref{inv}) corresponds to
\bea
|\vf\rangle\to \rho  e^{ie\alpha(\bs{X})} \rho^{-1}|\vf\rangle =e^{ie\alpha(\bs{x})}|\vf\rangle.
\eea

Since we are using $\bs{X}$ in the exponent in Eq.~(\ref{inv}), we will also need to
write $H$ in terms of $\bs{X}$ and the
corresponding conjugate observable $\bs{P}$, according to Eq.~(\ref{H equivalence}), i.e.
\bea
H(\bs{x},\bs{p})=h(\bs{X},\bs{P}).
\eea
The minimal substitution we require, in $h(\bs{X},\bs{P})$, is then
\bea\label{minsub}
\bs{P}\to \bs{P}-e\bs{A}(\bs{X})
\eea
with the combined transformations
\bea\label{GT}
\left\{\begin{array}{l}|\psi\rangle\to |\psi'\rangle=e^{ie\alpha(\bs{X})}|\psi\rangle\\ \\
\ \ \bs{A}(\bs{X})\to \bs{A}'(\bs{X})=\bs{A}(\bs{X})-\nabla_{\bs{X}}\alpha(\bs{X})\ .\end{array}\right.
\eea
It is important to note that because $\bs{X}$ and $\bs{x}$ do not commute, the argument of $\bs{A}$ in
Eq.~(\ref{minsub}) must be $\bs{X}$ rather than $\bs{x}$ in order
to ensure that
$$
e^{-ie\alpha(\bs{X})} (\bs{P}-e \bs{A}')e^{ie\alpha(\bs{X})}=\bs{P}-e \bs{A}.
$$

Given the gauge transformation of Eq.~(\ref{GT}), we are obliged to define $\bs{B}(\bs{X})=\nabla_{\bs{X}} \times\bs{A}(\bs{X})$,
and the Fourier transform of Eq.~(\ref{FT}) will also have to be rewritten in terms of $\bs{X}$.
How are we to interpret this, when $\bs{X}$ is a complicated non-local operator?
The answer is that the external, classical electromagnetic potential is in reality $\bs{A}(\bs{\xi})$, where $\bs{\xi}$
is a real vector of position. Then $\bs{B}(\bs{\xi})=\nabla_{\bs{\xi}} \times\bs{A}(\bs{\xi})$, and Eq.~(\ref{FT}) becomes
\bea\label{FTp}
\bs{A}(\bs{\xi},t)=\int d\omega \tilde{\bs{A}}(\omega) e^{i(\bs{k.\xi}-\omega t)} + c.c.
\eea
Then, in the interaction with the non-Hermitian system, $\bs{\xi}$ is replaced by the operator $\bs{X}$,
of which it is the eigenvalue. This is in parallel with the normal practice whereby in Eq.~(\ref{FT}) it is understood that
$\bs{x}$ is a numerical vector, but in its interaction with a Hermitian system $\bs{x}$ is interpreted as the operator
$\bs{\hat{x}}$.

If $h$ is of standard form, $\bs{p}^2/(2\mu)+U(x)$, the scattering rate is
\bea
w_{ij}&\propto& \frac{e^2}{\mu^2}|\langle \psi_i|\eta P_A|\psi_j\rangle|^2\nonumber\\
&=&\frac{e^2}{\mu^2}|\langle \vf_i|p_A|\vf_j\rangle|^2,
\eea
and the second form of the matrix element can then be rewritten, as in the Hermitian case, as a matrix element of $x_A$, namely
\bea
\langle \vf_i|p_A|\vf_j\rangle=i\mu \omega_{ij}\langle \vf_i|x_A|\vf_j\rangle.
\eea

\subsection{The Swanson model}
\noindent A much-studied example where $h$, but not $H$, is of standard form is the Swanson Hamiltonian\cite{Swanson},
whose three-dimensional version reads
\bea
H=\frac{\bs{p}^2}{2m_1}+\half i \omega \ve \{x_r, p_r\} +\half m_2\omega^2\bs{x}^2\ ,
\eea
with $m_2=(1-\ve^2)m_1$. There is actually a one-parameter family\cite{HG} of $Q$s, from which we consider
just the two cases (i) $Q=Q(\bs{x})$ and (ii) $Q=Q(\bs{p})$. In either case the equivalent Hermitian
Hamiltonian is just a harmonic oscillator of the form
\bea
h(\bs{x},\bs{p})=\frac{\bs{p^2}}{2\mu}+\half \mu \omega^2 \bs{x}^2
\eea

\subsubsection*{(i)\ \ \ $Q=Q(\bs{x})=\ve\,m_1\omega\,\bs{x}^2.$}
This amounts to completing the square as
\bea
H=\frac{(\bs{p}+i\ve\,m_1\omega\bs{x})^2}{2m_1}+\half m_1\omega^2\bs{x}^2
\eea
so that $\bs{X}=\bs{x}$, while $\bs{P}=\bs{p}+i\ve\,m_1\omega\bs{x}$. Thus in this case
\bea
h(\bs{x},\bs{p})=\frac{\bs{p}^2}{2m_1}+\half m_1\omega^2\bs{x}^2,
\eea
so that $\mu=m_1$. The coupling to the vector potential is thus
\bea
-\frac{e}{2m_1}(\bs{A}.\bs{P}+\bs{P}.\bs{A})
=-\frac{e}{2m_1}\left[(\bs{A}.\bs{p}+\bs{p}.\bs{A})+i\ve\,m_1\omega(\bs{A}.\bs{x}+\bs{x}.\bs{A})\right]
\eea
The required matrix element
\bea\label{ME}
\langle\psi_i|\eta P_A|\psi_j\rangle=\langle\vf_i|p_A|\vf_j\rangle,
\eea
is then found from expressing each component of $p$ on the right-hand side in terms
of creation and annihilation operators: $p=i\surd(m_1\omega/2)(a^\dag-a)$.
\subsubsection*{(ii) \ \  $Q=Q(\bs{p})=-\ve\,\bs{x}^2/(m_2\omega)$}

This amounts to completing the square instead as
\bea
H&=&\frac{\bs{p}^2}{2m_2}+\half m_2\omega^2\left(\bs{x}+\frac{i\ve\bs{p}}{m_2\omega}\right)^2\nonumber\\&&\\
&\equiv&\frac{\bs{P}^2}{2m_2}+\half m_2\omega^2\bs{X}^2\nonumber\ ,
\eea
so that $\bs{P}=\bs{p}$, while $\bs{X}=\bs{x}+i\ve\bs{p}/(m_2\omega )$. Thus in this case
\bea
h(\bs{x},\bs{p})=\frac{\bs{p}^2}{2m_2}+\half m_2\omega^2\bs{x}^2,
\eea
with $\mu=m_2$. The coupling to the vector potential is thus
\bea
-\frac{e}{2m_2}(\bs{A}.\bs{P}+\bs{P}.\bs{A})
=-\frac{e}{2m_2}(\bs{A}.\bs{p}+\bs{p}.\bs{A})
\eea
The matrix elements are still of the form of Eq.~(\ref{ME}), but now
the components of $p$ on the right-hand side are expressed as $p=i\surd(m_2\omega/2)(a^\dag-a)$.

The important thing to notice is that one will get different transition rates in the two cases.
That is, the system is determined not only by the Hamiltonian $H$, but also by the particular
metric operator $\eta$ used to restore unitarity.

\subsection{Imaginary cubic interaction}
The more common situation is that $H$ is of standard form,
while $h$ is a complicated non-local object. For example, in the case of the (one-dimensional)
prototype Hamiltonian
\bea
H=\half(p^2+x^2)+igx^3,
\eea
we have\cite{BBJ-ix3}
\bea
Q = -g\left(\frac{4}{3}p^3 + 2xpx\right)+O(g^3),
\eea
which gives rise\cite{HFJ-ix3, AM-ix3} to the observables
\begin{equation}
\left.\begin{array}{l} X=x+ig(x^2+2p^2) +g^2(-x^3+2pxp)\\ \\
P=p-ig(xp+px)+g^2(2p^3-xpx)\end{array}\right\}+O(g^3).
\end{equation}
Referring to Eq.~(\ref{H equivalence}), we can write $H(x,p)$ as $h(X,P)$,
where $h(x,p)$ has been calculated up to second order in $g$ as\cite{HFJ-ix3, AM-ix3}
\begin{eqnarray}\label{h2}
h(x,p)&=&\half(p^2+x^2)+3g^2\left(\half x^4+S_{2,2}(x,p) -\frac{1}{6}\right)+O(g^4),
\end{eqnarray}
where $S_{2,2}(x,p)=(x^2p^2+xp^2x+p^2x^2)/3$.

From Eq.~(\ref{h2}), we see that the minimal substitution $P\to P-eA(X)$ in $h(X,P)$
will give rise to additional couplings, of order $g^2$, arising from the mixed term $S_{2,2}(X,P)$.

To $O(g)$ the matrix elements will be just $\langle \psi_i|\eta P_A|\psi_j\rangle$.
In order to calculate this we will need the
corrected eigenfunctions, which have a first-order contribution, namely
\bea
\psi_i(x)=\psi_i^0(x)+g\sum_{j\ne i}\langle \psi_j^0|ix^3|\psi_i^0\rangle \psi_j^0(x)+O(g^2)
\eea
In this case it is much easier\cite{CMBh} to work with $H$ directly rather than with $h$.

\section{Summary}
For a standard Hermitian system the coupling to the electromagnetic potential, via the minimal substitution
 $\bs{p}\to\bs{p}-e\bs{A}(\bs{x})$, is
induced by implementing the position-dependent phase change $\psi\to e^{ie\alpha(\bs{x})}\psi$ and demanding that the
transformed Schr\"odinger equation be unchanged. For a quasi-Hermitian system we find instead that the phase must be
taken as $\alpha(\bs{X})$,
where $\bs{X}$ is the observable associated with $\bs{x}$. The coupling to the electromagnetic vector potential
thus induced is via the minimal substitution $\bs{P}\to\bs{P}-e\bs{A}(\bs{X})$ in $H(\bs{x},\bs{p})$ written in terms of
$\bs{X}$ and $\bs{P}$, where $\bs{P}$ is the observable associated with $\bs{p}$.

The matrix elements governing electromagnetic transitions from one state of the system to another depend on both $H$
and the metric $\eta$. In the special case of the Swanson Hamiltonian, when the equivalent Hermitian Hamiltonian $h$ is local,
this dependence is encoded in the mass of the particle, which cannot simply be read off from $H$. Generically $h$ is not
local, and the coupling is considerably more complicated.\\

\noindent{\bf Acknowledgement}

\noindent I am grateful to the referee for pointing out a serious mistake in the first version of this paper.

\end{document}